# Multimode optical fiber based spectrometers

**Brandon Redding, Sebastien M. Popoff, Hui Cao**[*]

*Department of Applied Physics, Yale University, New Haven, CT 06520*
[*]*hui.cao@yale.edu*

**Abstract:** A standard multimode optical fiber can be used as a general purpose spectrometer after calibrating the wavelength dependent speckle patterns produced by interference between the guided modes of the fiber. A transmission matrix was used to store the calibration data and a robust algorithm was developed to reconstruct an arbitrary input spectrum in the presence of experimental noise. We demonstrate that a 20 meter long fiber can resolve two laser lines separated by only 8 pm. At the other extreme, we show that a 2 centimeter long fiber can measure a broadband continuous spectrum generated from a supercontinuum source. We investigate the effect of the fiber geometry on the spectral resolution and bandwidth, and also discuss the additional limitation on the bandwidth imposed by speckle contrast reduction when measuring dense spectra. Finally, we demonstrate a method to reduce the spectrum reconstruction error and increase the bandwidth by separately imaging the speckle patterns of orthogonal polarizations. The multimode fiber spectrometer is compact, lightweight, low cost, and provides high resolution with low loss.

## 1. Introduction

While traditional spectrometers are based on one-to-one spectral-to-spatial mapping, spectrometers can also operate on more complex spectral-to-spatial mapping [1-3]. In these implementations, a transmission matrix is used to store the spatial intensity profile generated by different input wavelengths. A reconstruction algorithm allows an arbitrary input spectrum to be recovered from the measured spatial intensity distribution. While this approach is more complicated than the traditional grating or prism based spectrometers, it affords more flexibility in the choice of dispersive element. For instance, spectrometers based on this approach have been demonstrated using a disordered photonic crystal [1], a random scattering medium [2], and an array of Bragg fibers [3]. We recently found that a multimode optical fiber is an ideal dispersive element for this type of spectrometer because the long propagation length and the minimal loss enables high spectral resolution and good sensitivity [4].

In a multimode fiber spectrometer, the interference between the guided modes creates a wavelength-dependent speckle pattern, providing the required spectral-to-spatial mapping. In the past, the contrast of this speckle pattern was found to depend on the spectral width and shape of the optical source [5-8], allowing researchers to use contrast as a measure of the laser linewidth [9]. As opposed to using only the statistical property of the speckle such as the contrast, we recently proposed and demonstrated that by recording the entire speckle patterns at different wavelengths, a multimode fiber can be used as a general purpose spectrometer [4]. The spectral resolution of the device depends on the spectral correlation width of the speckle, which is known to scale inversely with the length of the fiber [4,7,9]. The advantage of using an optical fiber is that a long propagation length is easily achieved with minimal loss, giving high spectral resolution. Furthermore, the fiber-based spectrometer requires only a multimode fiber and a monochrome CCD camera to record the speckle patterns. Compared to traditional spectrometers, optical fibers are lower cost, lighter weight, and can be coiled into a small volume while providing spectral resolution that is competitive with state-of-the-art grating-based spectrometers.

In this paper, we extend on the proof-of-concept demonstration presented in Ref. [4] and explore the operational limits of a multimode fiber spectrometer. We provide a theoretical analysis of the effects of the fiber geometry on the spectrometer performance, and then present a reconstruction algorithm combining a truncated inversion technique with a least squares minimization procedure, which enables accurate and robust spectral reconstruction in the presence of experimental noise. We also investigate the effects of spectral and spatial oversampling on the quality of the recovered spectra. Using a 20 meter long fiber, we are able to resolve two laser lines separated by merely 8 pm. A higher spectral resolution is expected for a longer fiber, but we are currently limited by the resolution of the tunable laser source used for calibration. We also discuss the bandwidth limitation when measuring a dense spectra due to speckle contrast reduction. To reduce the reconstruction error and increase the spectral bandwidth, we develop a method based on a polarization-resolved speckle measurement. Finally, we use a 2 centimeter long fiber to measure a continuous broadband spectrum generated by a supercontinuum source.

## 2. Operation Principle of Fiber Spectrometer

The fiber-based spectrometer consists of a multimode fiber and a monochrome CCD camera that images the speckle pattern at the end of the fiber. The speckle pattern, created by interference among the guided modes in the fiber, is distinct for light at different wavelength, thus providing a fingerprint of the input wavelength. In our experiments, we used commercially available step-index multimode fibers with 105 μm diameter cores (NA = 0.22) and lengths varying from 2 cm to 20 m. A schematic of the experimental configuration is shown in Fig. 1(a). A near-IR tunable diode laser (HP 8168F) was used to provide a spectrally controlled input signal for the calibration and initial characterization. A polarization maintaining single-mode fiber was used to couple the laser emission into the multimode fiber through a standard FC/PC mating sleeve. The speckle pattern generated at the exit face of the multimode fiber was imaged onto an InGaAs camera (Xenics Xeva 1.7-320) with a 50× near-IR microscope objective (NA=0.55). Alternatively, the speckle in the far-field zone may be projected directly onto the camera without the objective.

Figure 1(b) is a movie showing the speckle patterns recorded at the end of a 20 m fiber as the input wavelength is scanned from 1500 nm to 1501 nm in the step of 0.01 nm. The speckle patterns decorrelate for very small changes in wavelength. Such high spatial-spectral diversity gives fine spectral resolution. We calculated the spectral correlation function of the speckle intensity, $C(\Delta\lambda, x) = \langle I(\lambda, x) I(\lambda+\Delta\lambda, x) \rangle / [\langle I(\lambda, x) \rangle \langle I(\lambda+\Delta\lambda, x) \rangle] - 1$, where $I(\lambda, x)$ is the intensity at a position $x$ for input wavelength $\lambda$, $\langle ... \rangle$ represents the average over $\lambda$. In Fig. 1(c), we plot the spectral correlation function averaged over many spatial positions across the 105 μm core of a 20 m fiber. The spectral correlation width, $\delta\lambda$ is defined as $C(\delta\lambda/2) = C(0)/2$. When the input wavelength shifts by $\delta\lambda$, the output speckle pattern becomes nearly uncorrelated, $C(\Delta\lambda=\delta\lambda) \approx 0$. In this case, $\delta\lambda = 10$ pm.

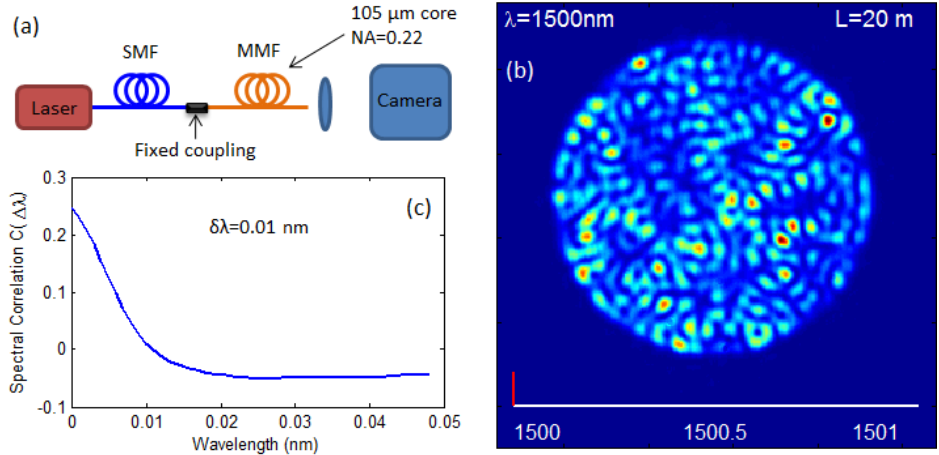

Fig. 1. (a) A schematic of the fiber spectrometer setup. A near-IR laser, wavelength tunable from 1435 nm to 1590 nm, is used for calibration and testing. Emission from the laser is coupled via a single-mode polarization-maintaining fiber (SMF) to the multimode fiber (MMF), with a standard FC/PC mating sleeve. A 50× objective lens is used to image the speckle pattern generated at the end facet of the fiber to the monochrome CCD camera. (b) (Multimedia) Movie showing the speckle pattern generated at the end of a 20 m multimode fiber as the input wavelength varies from 1500 nm to 1501 nm in the step of 0.01 nm. The wavelength is written on the top and also marked by the red line in the bottom scale. The speckle pattern decorrelates rapidly with wavelength, illustrating high spatial-spectral diversity. (c) Spectral correlation function of speckle intensity obtained from (b) exhibits a correlation width $\delta\lambda = 0.01$ nm, which enables fine spectral resolution.

To use the fiber as a spectrometer, speckle patterns such as the ones shown in Fig. 1(b) are recorded to construct the transmission matrix, as will be discussed in detail in section 4. After this calibration step, the tunable laser can be replaced by any optical source and the camera will record the speckle pattern. A reconstruction algorithm, to be discussed in section 4, is then applied to recover the spectra of the input. Note that in the experiments presented in this work, a single-mode polarization-maintaining fiber is always used to couple the signal to the multimode fiber. This ensures that the input to the multimode fiber will have the same spatial profile and polarization as the calibration. If the probe signal had a different profile or polarization, it could excite a different combination of fiber modes with different (relative) amplitudes and phases, making the calibration invalid.

## 3. Effects of fiber geometry on spectral correlation of speckle

In this section, we present a theoretical analysis of the dependence of the spectrometer resolution on the fiber geometry. The fiber length $L$, the core diameter $W$, and the numerical aperture NA are crucial parameters determining the resolution and bandwidth of the fiber-based spectrometer. If we consider a monochromatic input light which excites all the guided modes, then the electric field at the end of a waveguide of length $L$ is:

$$\mathbf{E}(r,\theta,\lambda,L) = \sum_m A_m \mathbf{\psi_m}(r,\theta,\lambda) \exp\left[-i\left(\beta_m(\lambda)L - \omega t + \varphi_m\right)\right] \quad (1)$$

where $A_m$ and $\varphi_m$ are the amplitude and initial phase of the $m^{th}$ guided mode which has the spatial profile $\mathbf{\psi_m}$ and propagation constant $\beta_m$. To simplify the analysis, we considered planar waveguides of width $W$ and numerical aperture NA, and calculated the mode profile and propagation constant using the method outlined in Ref. [10]. We assumed that initially all of the modes in the waveguide were excited equally ($A_m=1$ for all modes) with uncorrelated phases ($\varphi_m$ are random numbers between 0 and $2\pi$). We calculated the spatial distribution of electric field intensity at the end of a waveguide, as a function of the input wavelength, and then computed the spectral correlation function of intensity $C(\Delta\lambda)$. By repeating this process as we varied $L$, $W$, or NA, we were able to obtain the dependence of the spectral correlation width $\delta\lambda$ on these parameters. Figure 2(a) plots $\delta\lambda$ vs. $L$ for $W = 1$ mm and NA = 0.22. $\delta\lambda$ scales inversely with $L$, indicating that the spectral resolution can be changed simply by varying the waveguide length. Next we fix $L = 1$ m, and plot $\delta\lambda$ as a function of $W$ in Fig. 2(b) for NA = 0.22. The spectral correlation width initially drops, but quickly saturates. In Fig. 2(c), we varied NA while keeping $L = 1$ m and $W = 1$ mm, and observed a reduction of $\delta\lambda$ with NA.

To obtain a physical understanding of the simulation results, we present a qualitative analysis of speckle decorrelation using Eq. 1. Individual modes propagate down the waveguide with different propagation constants ($\beta_m$) and accumulate different phase decays ($\beta_m L$). The maximum difference occurs between the fundamental mode ($m = 1$) and the highest-order mode ($m = M$), which we denote $\varphi(\lambda) = \beta_1(\lambda)L - \beta_M(\lambda)L$. In order for an input wavelength shift $\delta\lambda$ to produce a distinct intensity distribution at the output, $\varphi(\lambda)$ should change by approximately $\pi$, namely, $|d\varphi(\lambda)/d\lambda|\ \delta\lambda \sim \pi$. By approximating $\beta_1(\lambda) \approx k$ and $\beta_M(\lambda) \approx k\cos(\text{NA})$ for large $W$ ($k = 2\pi n/\lambda$, $n$ is the refractive index of the waveguide, $\lambda$ is the vacuum wavelength), we get $\delta\lambda \sim (\lambda/n)^2/(2\ n\ L)/[1 - \cos(\text{NA})]$. For small NA, $1 - \cos(\text{NA}) \approx (\text{NA})^2/2$, and $\delta\lambda \sim \lambda^2 / [n\ L\ (\text{NA})^2]$. This simple expression captures the basic trends in Fig. 2(a-c). In Fig. 2(a), the spectral correlation width scales linearly with $1/L$. When $W$ is large, the spectral decorrelation does not depend on $W$ [Fig. 2(b)]. Figure 2(c) is a log-log plot showing the calculated $\delta\lambda$ fit well with a straight line of slope -2.09, thus confirming the $1/(\text{NA})^2$ scaling.

The above analysis assumed that the spectral correlation width was determined by the total range of the propagation constants in the waveguide, $\beta_1$ - $\beta_M$. This assumption led to the strong dependence of $\delta\lambda$ on NA, which dictates the highest order mode the waveguide supports. The waveguide width $W$, which predominantly determines the $\beta$ spacing of adjacent modes, had little effect on $\delta\lambda$. To validate this assumption, we considered a waveguide with fixed geometry ($L$=1 m, $W$=1 mm, and NA=0.22) and calculated the spectral correlation function when exciting a subset of the ~300 guided modes (i.e. $A_m = 0$ for some $m$'s). Figure 2(d) plots $C(\Delta\lambda)$ in four cases: (i) all the modes are excited; (ii) only the 100 lowest order modes are excited; (iii) only the 100 highest order modes are excited; and (iv) every 5th mode is excited. The spectral correlation width decreased in (ii) and (iii), but remained virtually unchanged in (iv). These results confirm that the speckle decorrelation depends primarily on the difference between the maximal and minimal $\beta$ values. Increasing $W$ adds modes in between $\beta_1$ and $\beta_M$, but does not significantly modify the values of $\beta_1$ and $\beta_M$, thus having little impact on the spectral resolution. However, as will be discussed later, increasing W does increase the spectral bandwidth of the fiber spectrometer.

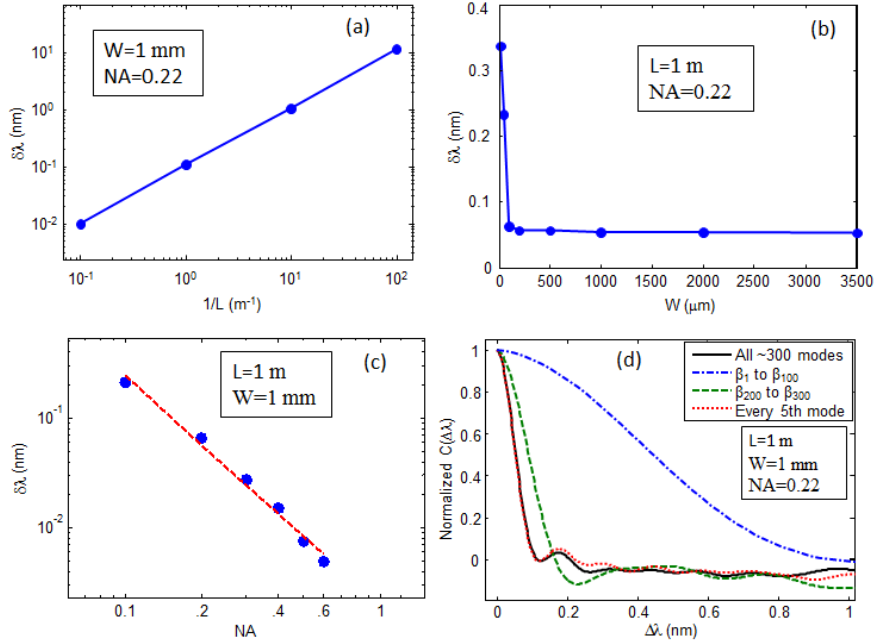

Fig. 2. (a) Calculated spectral correlation width $\delta\lambda$ as a function of the length $L$ of a planar waveguide with a fixed width $W$=1 mm and NA=0.22. The wavelength of the input light is 1500 nm. $\delta\lambda$ scales linearly with $1/L$ (blue line), indicating that a longer waveguide provides finer spectral resolution. (b) Calculated spectral correlation width $\delta\lambda$ as a function of the width $W$ of a planar waveguide with a fixed length $L$=1 m and NA=0.22. The wavelength of the input light is 1500 nm. $\delta\lambda$ decreases quickly at small $W$ and then saturates at large $W$. (c) Calculated spectral correlation width $\delta\lambda$ as a function of the numerical aperture NA of a planar waveguide with $L$=1 m and $W$=1 mm. The wavelength of the input light is 1500 nm. Red dotted line is a linear fit in the log-log plot of the calculated $\delta\lambda$ (blue dots) vs. NA, it has a slope of -2.09, indicating $\delta\lambda$ scales as $1/NA^2$. (d) Calculated spectral correlation function of the intensity at the end of a planar waveguide ($L$=1 m, $W$=1 mm, NA = 0.22), when a subset of the waveguide modes were excited. The waveguide supports ~300 modes at $\lambda$ = 1500 nm. (i) All the modes were excited (black solid line), (ii) only the 100 lowest order modes were excited (blue dash-dotted line), (iii) only the 100 highest order modes were excited (green dashed line), (iv) every fifth mode was excited (red dotted line). The spectral correlation width decreased in the cases of (ii) and (iii), but remained virtually unchanged in (iv), indicating the speckle decorrelation is determined primarily by the total range of $\beta$ values of the modes that are excited by the input signal.

Finally, we discuss the experimental implication of the above analysis. Figure 2(d) reveals that for a given fiber the fastest speckle decorrelation is reached only when we excite the guided modes that cover the full range of the propagation constants. Experimentally, the measured spectral correlation width of a given fiber changed by as much as ~50% depending on the coupling of the input, which could be varied by adjusting the lateral alignment of the single-mode fiber to the multimode fiber. In the future, chaotic multimode fibers might be used to avoid the alignment issue and ensure all of the modes are excited more or less equally [11]. The non-circular cross-section of the core, e.g. the D-shaped cross-section, leads to chaotic dynamics of light rays in the fiber. Consequently the majority of the guided modes are spread uniformly over the entire core, and they will be excited no matter where light enters the fiber at the input facet.

**4. Algorithm for spectral reconstruction**

In the previous section, we investigated the effect of the fiber geometry on speckle decorrelation, and argued that the spectral correlation width determines the spectral resolution. More accurately, the spectral correlation provides a limit on the resolution, since we cannot distinguish between two wavelengths that produce highly correlated speckle patterns. In practice, due to the presence of experimental noise, the resolution of the fiber-based spectrometer also relies on the algorithm used to reconstruct the input spectrum from a measured speckle pattern. In this section, we describe a robust algorithm capable of reconstructing an arbitrary input spectrum in the presence of noise.

Although the speckle patterns generated by different wavelengths could theoretically be calculated, as discussed in the previous section, this is not practical for a real fiber because it would require precise knowledge of the fiber geometry, including any twisting or bending, as well as the spatial profile of the input signal. Instead, we experimentally calibrated the transmission matrix for an input signal with a fixed spatial profile and polarization. The fiber was coiled and secured to an optics table before calibration and was not moved during subsequent testing. The monochrome CCD camera captures the intensity distribution across the exit face of the fiber, $I(\mathbf{r}) = \int S(\lambda) F(\mathbf{r},\lambda) A(\lambda) d\lambda$ where $S(\lambda)$ is the spectral flux density of the input signal, $F(\mathbf{r}, \lambda)$ is the position dependent transmission function of the fiber, $A(\lambda)$ is the spectral sensitivity of the camera. Instead of measuring the spectral responses of the fiber and the camera separately, our calibration procedure measures the total transmission function of the fiber spectrometer, defined as $T(\mathbf{r},\lambda) = F(\mathbf{r},\lambda) A(\lambda)$. It characterizes the spectral to spatial mapping from the input to the output of the fiber, as well as the spectral response of the camera:

$$I(\mathbf{r}) = \int S(\lambda) T(\mathbf{r},\lambda) d\lambda \qquad (2)$$

In practice, the spectral signal is discretized into spectral channels centered at $\lambda_i$ and spaced by $d\lambda$. If adjacent spectral channels are separated by more than the spectral correlation width $\delta\lambda$, they become independent because their speckle patterns are uncorrelated. Similarly, spatial discretization across the speckle image generates the spatial channels centered at $r_j$, and they become independent if their spacing $dr$ exceeds the spatial correlation length $\delta r$ (equal to the average speckle size). In section 5, we will discuss this discretization process in more detail and explore the choice of the spatial and spectral channel spacing. In this section, we set $d\lambda = \delta\lambda/2$ and $dr = \delta r$. After discretizing the speckle patterns and the input spectrum, the transmission function $T$ becomes a discrete matrix, and Eq. (2) becomes

$$\mathbf{I} = \mathbf{T S}, \qquad (3)$$

where $\mathbf{I}$ is a vector representing the intensities in $N$ spatial channels of the output, and $\mathbf{S}$ the intensities in $M$ spectral channels from the input. Each column in the transmission matrix $\mathbf{T}$ describes the discretized speckle pattern, $I_r$, produced by incident light in one spectral channel.

As an example, we consider the transmission matrix corresponding to a 1 m long fiber with spectral correlation width $\delta\lambda = 0.4$ nm. We calibrated a transmission matrix from $\lambda = 1450$ nm to 1550 nm in steps of 0.2 nm providing $M = 500$ spectral channels. From the spatial correlation function, we found that the speckle image contained ~600 spatial channels separated by $\delta r$, and we sampled 500 of these, $N = 500$. In section 5, we will discuss these choices in more detail. The transmission matrix was calibrated one column at a time by recording the speckle pattern generated at each sampled wavelength $\lambda_i$ in $\mathbf{S}$. The entire calibration process, consisting of recording speckle images at 500 input wavelengths, was completed in a few minutes. After calibration, we tested the spectrometer operation by measuring the speckle pattern produced by a probe signal and attempted to reconstruct the probe spectrum. An initial test was to recover a spectrum consisting of three narrow lines with unequal spacing and varying height. Since optical signals at different wavelengths do not interfere, we were able to synthesize the probe speckle pattern by adding weighted speckle patterns measured sequentially at the three probe wavelengths.

In our first attempt to reconstruct the input spectrum, we simply multiplied the measured speckle pattern by the inverse of the transmission matrix: $S = T^{-1} I$. In Fig. 3(a), we plot the original spectrum with a red dotted line and the reconstructed spectrum with a blue solid line. This simple approach failed because the inversion of matrix $T$ is ill-conditioned in the presence of experimental noise. There are several sources of noise in our spectrometer: (i) discrete intensity resolution of a CCD array, (ii) ambient light, (iii) mechanical instability of the fiber and fluctuation of experimental environment, (iv) intensity fluctuation and wavelength drift of the laser source. Among them we believe (iii) has the dominant contribution.

To understand why even a small amount of experimental noise can corrupt the reconstruction, we employ the singular value decomposition to factorize the transmission matrix, $T = U D V^T$, where $U$ is a $N \times N$ unitary matrix, $D$ is a $N \times M$ diagonal matrix with positive real elements $D_{jj}=d_j$, known as the singular values of $T$, $V$ is a $M \times M$ unitary matrix. The rows of $V$ (resp. columns of $U$) are the input (resp. output) singular vectors and are noted $V_j$ (resp. $U_j$). To find the inverse of $T$, we take the reciprocal of each diagonal element of $D$ and then transpose it to obtain a diagonal matrix $D'$. The inverse of $T$ is given as: $T^{-1}=VD'U^T$. The issue we confront in the presence of noise is that the small elements of $D$ are the elements most corrupted by experimental noise, and they are effectively amplified in $D'$ when we take their reciprocal. To overcome this issue, we adopted a "truncated inversion" technique, similar to the approach in Ref. [3]. In this approach, we compute a truncated version of $D'$ in which we only take the reciprocal of the elements of $D$ above a threshold value and set the remaining elements to zero. The truncated inverse of $T$ is $T_{trunc}^{-1} = V D'_{trunc} U^T$, which we use to reconstruct the input spectrum $S = T_{trunc}^{-1} I$. Using this truncated inversion technique, we obtained a far superior reconstruction of the input spectrum, as shown in Fig. 3(b). Clearly the truncated inversion technique was able to recover the input spectrum. To provide a quantitative measure of the quality of spectral reconstruction, we calculated the spectrum reconstruction error, defined as the standard deviation between the probe spectrum and the reconstructed spectrum: $\mu = \sqrt{\frac{1}{M}\sum_\lambda \left[S_{probe}(\lambda) - S_{reconstructed}(\lambda)\right]^2} \bigg/ \left[\frac{1}{M}\sum_\lambda S_{probe}(\lambda)\right]$. In order to optimize the truncated inversion, we reconstructed the input spectra using different threshold values for truncation and calculated the reconstruction error $\mu$. The threshold of truncation was defined as a fraction of the largest element in $D$. For example, at a threshold of 0.01, any element in $D$ with amplitude less than 1% of the maximal element in $D$ would be discarded in the inversion process. In Fig. 3(c), we show the spectrum reconstruction error as a function of the truncation threshold. We found that the optimal threshold value was $\sim 2 \times 10^{-3}$ [this threshold value was used to reconstruct the spectrum shown in Fig. 3(b)]. After optimizing the truncation threshold, the singular value decomposition is only performed once for a given fiber, providing a $T_{trunc}^{-1}$ matrix which can then be used to recover any input spectrum with a single matrix multiplication.

The threshold is directly related to the experimental noise, when the noise increases, more singular values are perturbed and have to be discarded. One can understand how the optimal threshold behaves in the presence of noise by looking at its effect on the singular vectors of the $T$. Consider the perturbed matrix $T_p = T + G$, where $G$ is a matrix of Gaussian noise with standard deviation $\sigma_{exp}$, representing the experimental error and $T = U D V^T$ is the transmission matrix of the system without noise. We can write $T_p V_j = d_j U_j + G V_j$, with $d_j$ the $j^{th}$ singular value of $T$ and $G V_j$ a perturbation term. We have $\|d_j U_j\| = d_j$ and $\langle \|G V_j\| \rangle = \sigma_{exp} \sqrt{M}$, where $\langle ... \rangle$ indicates averaging over realizations of the noise. We want to discard the singular values lower than the noise level, i.e. the singular values following $d_j < \sigma_{exp} \sqrt{M}$. To obtain the threshold as defined previously, we have to compare this value to the largest singular value of

$T$, which can easily be estimated when $T$ is composed of independent elements, identically distributed with a mean value not equal to zero by $\langle d_1 \rangle = \tau\sqrt{NM}$, with $\tau$ the average value of the elements of $T$ (i.e. the mean speckle intensity). We finally obtain an optimal threshold of the order of $\sigma_{exp}N^{-1/2}\tau^{-1}$. In the reported experiment, we measured the noise variance by sequentially recording the matrix $T$ twice for the same system, yielding $\sigma_{exp} = 5\times10^{-3}$. Based on this noise variance, we predicted an optimal threshold value of $1.2\times10^{-3}$ in good agreement with the optimal value of $2\times10^{-3}$ observed experimentally.

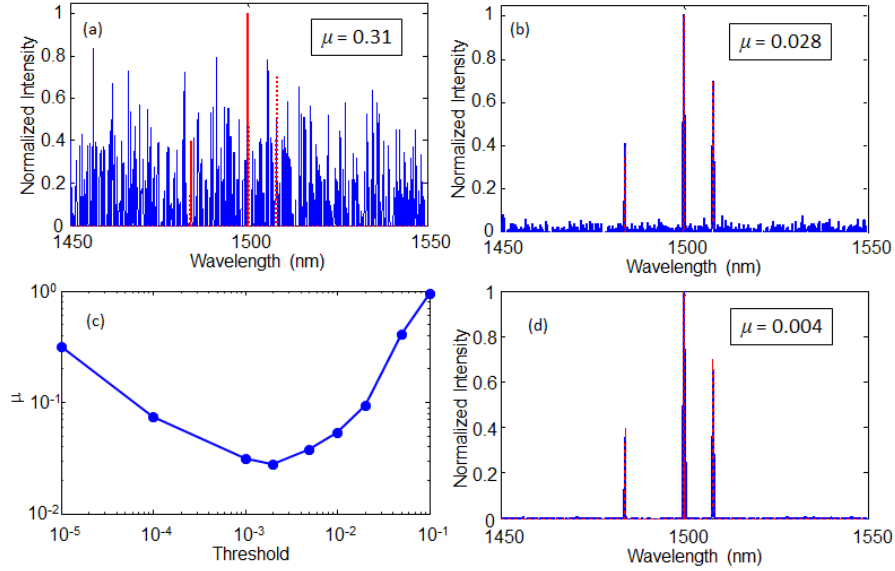

Fig. 3. (a) Initial attempt to reconstruct a probe spectrum, consisting of three narrow lines centered at 1484, 1500, and 1508 nm and with relative amplitude of 0.4, 1, and 0.6 (indicated by the red dotted line), simply by inversion of the transmission matrix: $S = T^{-1} I$ fails, because this inversion process is ill-conditioned in the presence of experimental noise. (b) The same probe spectrum reconstructed using the truncated inversion technique described in the text. The truncation threshold was set to $2\times10^{-3}$, and the spectrum reconstruction error $\mu = 0.028$. The truncated inversion technique was able to recover the input spectrum, although background noise is still evident. (c) The spectrum reconstruction error, $\mu$, of the reconstructed spectrum as a function of the truncation threshold. The minimal $\mu$ gives the optimal threshold value. (d) A further improved reconstruction was achieved using a nonlinear optimization procedure. The spectrum obtained from the truncated inversion technique was used as a starting guess to reduce the computation time, and the spectrum reconstruction error $\mu = 0.004$.

In order to further improve the quality of spectral reconstruction, we also developed a nonlinear optimization algorithm. The spectral reconstruction process can be framed as an energy minimization problem, in which the optimal solution of the input spectrum, $S$, is the one that minimizes the energy $E = \|I - T S\|^2$ [1]. To find the optimal $S$, we developed a simulated annealing algorithm. At each step in the optimization routine, we changed one element in $S$ at a time by multiplying it by a random number between 0.5 and 2. This provided us with a new spectrum, $S'$. We then calculated the change in energy $\Delta E = \|I - T S'\|^2 - \|I - T S\|^2$ and kept the change (i.e. $S = S'$) with a probability equal to $\exp[-\Delta E/T_0]$, where $T_0$ is the "temperature". After performing this process once for every element in $S$, the temperature was reduced and the process was repeated. At high temperature the algorithm is more likely to accept "bad" choices (which increase the energy) in order to search broadly for a global minimum. Later in the algorithm, at lower temperatures, the algorithm is less likely to accept a "bad" choice. Note that changes which reduce the energy ($\Delta E < 0$) are always accepted. The algorithm stopped after a fixed number of steps or if the energy dropped below a threshold value. In Fig. 3(d), we show the reconstructed spectrum obtained using the simulated annealing algorithm. The background noise is largely suppressed and the spectrum reconstruction error is reduced. The simulated annealing algorithm typically required a few hundred iterations to reach the optimal solution; however, the truncated inversion technique provides a good initial guess of *S*, which dramatically reduced the simulation time.

In practice, these two reconstruction algorithms could be used in tandem. The truncated inversion technique provides a decent near-instantaneous reconstruction, which operates in real-time, since it requires a single matrix multiplication. The simulated annealing algorithm could be used to obtain a more accurate spectrum in applications where fast temporal response is less critical, or after the measurement is concluded. Of course, in both cases, the reconstruction will be improved by reducing the noise of the experimental measurements. In the current setup, the main source of noise is expected to be the stability of the multimode fiber, which was secured to an optical table during testing. Improved methods to rigidly stabilize the multimode fiber are expected to reduce the noise and provide more accurate spectral reconstructions.

Using the reconstruction algorithm that combines truncated inversion and simulated annealing, we characterized the spectral resolution of the fiber spectrometer by testing its ability to discriminate between two closely spaced spectral lines. In order to synthesize the probe speckle pattern, we separately recorded speckle patterns at the two probe wavelengths and then added them in intensity. In Fig. 4, we show the reconstructed spectrum measured using a 20 m long fiber. The spectral positions of the probe lines (red dotted line) are in between the sampled wavelengths $\lambda_i$ used in calibration. The reconstructed spectrum (blue solid line) clearly identifies the two peaks even though they are separated by merely 8 pm. Even higher spectral resolution is possible by using a longer fiber; however we would need a tunable laser with finer spectral resolution in order to calibrate such a fiber. This level of spectral resolution is competitive with the top commercially available grating based spectrometers.

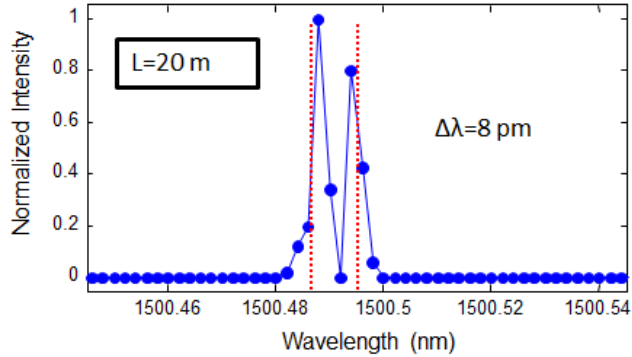

Fig. 4. Reconstructed spectrum (blue solid line with circular dots marking the calibrated wavelengths $\lambda_i$) of two spectral lines separated by 8 pm. It is obtained with a 20 m long fiber spectrometer. The red dotted lines mark the center wavelengths of the input lines.

## 5. Spatial and spectral sampling

In contrast to the conventional spectrometers which map spectral information to one spatial dimension, our fiber spectrometer maps to two-dimensions (2D). This 2D spatial-spectral mapping fully utilizes the large detection area of modern 2D cameras to achieve large bandwidth of operation. The maximal number of independent spectral channels that can be measured in parallel is limited by the number of independent spatial channels that encode the spectral information. The number of independent spatial channels is equal to the number of speckles in the intensity distribution at the output end of the fiber, and it is also equal to the total number of guided modes, *B,* in the fiber if all modes are more or less equally excited. *B* is determined by the fiber core diameter, $W$, the NA, and the wavelength of light $\lambda$, $B \approx \pi^2 (NA)^2 W^2 / 2\lambda^2$ [12]. The fibers considered in this work ($W$ = 105 μm, NA = 0.22) support ~1000 modes at an operating wavelength of 1500 nm. In principle, increasing $W$ or NA will increase the number of channels for transmitting spectral information, thus increasing the bandwidth of the fiber spectrometer. In reality, however, the experimental noise may reduce the number of spectral channels that can be recovered simultaneously due to degradation of the spectral reconstruction. In this section, we investigate how the spectrum reconstruction error is affected by spectral and spatial sampling rates.

We consider a 1 m long fiber with spectral correlation width $\delta\lambda$ = 0.4 nm and a spatial correlation width $\delta r$ = 4.1 μm. Our tunable laser covers the wavelength range from 1450 nm to 1550 nm, proving a maximum operating bandwidth of 100 nm. The speckle image had a radius of ~52.5 μm, providing ~600 spatial channels separated by $\delta r$. This number is less than the total number of guided modes in the fiber, because we did not excite all the modes [13]. Using this fiber, we generated various transmission matrices by adjusting the number of spectral and spatial channels, and the spacing of these channels. We then evaluated the reconstruction of a probe spectrum using different transmission matrices. The probe spectrum was a Lorentzian line centered at 1500 nm with a 1 nm full-width half-maximum (FWHM), synthesized by summing speckle patterns measured sequentially using the tunable laser, as discussed in the previous section. We used the truncated inversion algorithm to reconstruct the spectra and calculated the reconstruction error.

We first investigated the effect of the operating bandwidth on the ability of the spectrometer to reconstruct the Lorentzian spectrum. In this case, we fixed the spectral channel spacing at $\delta\lambda/2$ = 0.2 nm and the spatial channel spacing at $\delta r$ = 4.1 μm, and selected $N$ = 500 spatial channels. We then constructed transmission matrices using 125, 250, or 500 spectral channels, corresponding to an operating bandwidth ($\Delta\lambda = M \delta\lambda/2$) of 25, 50 or 100 nm. In Fig. 5(a), we present the spectrum reconstruction error $\mu$ for the three transmission matrices. We found that the error decreased with bandwidth. This is expected since less bandwidth (with fixed channel spacing) essentially means there are fewer spectral channels to recover. In other words, the number of variables ($M$) to solve for during the reconstruction is less than the number of equations given by the number of measured spatial channels ($N$). The redundant equations reduce the uncertainty in determining the intensity in each spectral channel. The larger the number of redundant equations ($N-M$), the lower the reconstruction error. Of course, reducing the bandwidth ($M$) also limits the utility of the spectrometer, so this may not always be practical. However, increasing the fiber core radius $W$ would increase the number of available spatial channels so that $M < N$ would still be possible even for a large bandwidth.

We then considered the effect of changing the spectral channel spacing for a transmission matrix with a fixed bandwidth. First, we set the bandwidth $\Delta\lambda$ = 100 nm and kept the same number of spatial channels $N$ = 500, but varied the spectral channel spacing $d\lambda$. We constructed matrices with $d\lambda$ = 0.2, 0.4, and 0.8 nm, corresponding to $M$=500, 250, and 125. As shown in Fig. 5(b), the spectrum reconstruction error decreased for larger spectral channel separation. Again, this corresponds to fewer spectral channels to recover. However, increasing the spectral channel spacing much beyond the spectral correlation width could lead to errors in the reconstruction. If the spectral channels are separated too far, an optical signal falling between the discretized wavelengths will generate a speckle pattern with minimal correlation to the speckle patterns stored in the transmission matrix, and its wavelength could not be recovered. For example, in Fig. 4 the probe lines were set at wavelengths in between the discretized wavelengths used in the transmission matrix. Since $d\lambda = \delta\lambda/2$, the speckle pattern at the probe wavelengths are still partially correlated with the speckle patterns generated at the nearby discretized wavelengths in $T$. Hence, the reconstruction algorithm is able to identify the wavelength of the probe signal. If the discretized spectral channels were separated by much more than the spectral correlation width, $d\lambda \gg \delta\lambda$, optical signals falling between these channels would have speckle patterns distinct from any speckle pattern recorded in the transmission matrix, thus recovery of their wavelengths becomes impossible.

Finally, we considered the effect of spatial oversampling. In this case, we used 500 fixed spectral channels separated by $d\lambda = \delta\lambda/2 = 0.2$ nm, providing 100 nm bandwidth. Since the speckle image contained ~600 speckles of average size $\delta r = 6$ pixel of the CCD camera and the total number of pixels in the image was ~17,000, there was plenty of room for oversampling. We therefore generated transmission matrices with 500, 1000, and 2000 spatial channels, corresponding to channel spacing $dr = 6$, 4, and 3 pixel. As shown in Fig. 5(c), increasing the number of spatial channels improves the spectral reconstruction. Even though the additional spatial channels are correlated, they provide redundancy which aids the reconstruction in the presence of noise. Furthermore, the number of available independent spatial channels can be increased by using a fiber with a larger core. As discussed in Section 2, the core diameter $W$ will not significantly affect the spectral resolution and could be adjusted to support a desired number of spatial channels. The downside to spatial oversampling is that the transmission matrix is larger, slowing down the reconstruction algorithm.

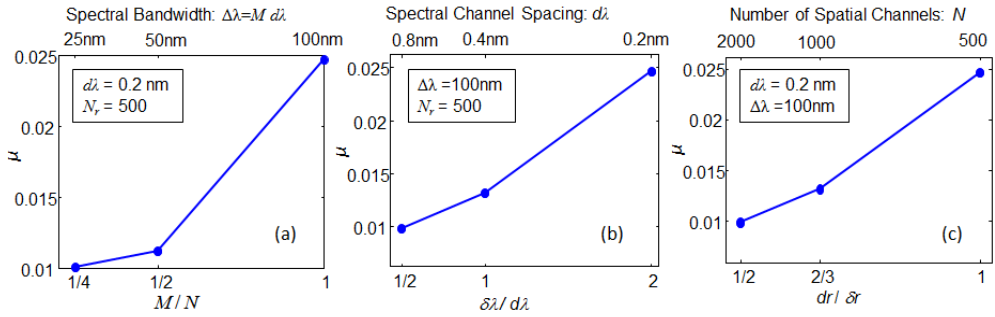

Fig. 5. The data points represent the spectrum reconstruction error, $\mu$, for a Lorentzian spectrum of FWHM = 1 nm using different transmission matrices, which were constructed for a 1 m long fiber with spectral correlation width $\delta\lambda = 0.4$ nm. (a) With a fixed spectral channel spacing $d\lambda = 0.2$ nm and $N = 500$ spatial channels, we compare the spectrum reconstruction error using the transmission matrices with varying bandwidth, $\Delta\lambda$. By reducing the bandwidth, the accuracy of the spectrum reconstruction is improved. (b) With a fixed 100 nm bandwidth and $N = 500$ spatial channels, we varied the spectral channel spacing, $d\lambda$. By increasing the channel spacing, the error of spectrum reconstruction reduced. (c) With a fixed bandwidth of 100 nm and spectral channel spacing of 0.2 nm, we varied the number of spatial channels extracted from the speckle pattern by reducing their spacing $dr$. Increasing the number of spatial samples, even though they were correlated ($dr < \delta r$), improved the accuracy of spectrum reconstruction.

## 6. Fiber-spectrometer bandwidth

Thus far, we have used the fiber spectrometer to reconstruct spectrally sparse, narrowband signals. If all the guided modes in the fiber are excited by the probe signal, the bandwidth for a sparse spectrum (most of the spectral channels carrying no signal) is $\Delta\lambda = B\,\delta\lambda$, where $B$ is the total number of guided modes that the fiber supports, and $\delta\lambda$ is the spectral correlation width of the speckle pattern. For a given fiber core diameter and numerical aperture, $B$ is fixed, while $\delta\lambda$ varies with the fiber length. A longer fiber gives better spectral resolution (smaller $\delta\lambda$), but narrower bandwidth. Such a tradeoff between the resolution and the bandwidth is similar to that of a traditional spectrometer. What is different from a traditional spectrometer is that the spectral coverage in a single measurement does not need to be continuous. If it is known *a priori* that the probe signal does not have components within certain spectral regions, such regions can be excluded from the spectrum reconstruction, thus the finite number of spectral channels ($B$) may cover an even broader spectral range.

Next, we investigate the ability of the fiber spectrometer to recover spectrally dense, broadband signals. The bandwidth for the dense spectra (in which optical signals are present at many spectral channels) is further limited by the speckle contrast. When the input signal is spectrally broad, different wavelengths produce distinct speckle patterns which add in intensity causing the speckle contrast to decrease. Since the speckle contrast is effectively the "signal" used to reconstruct the input spectrum, the reconstruction degrades as the contrast is reduced. In particular, as the speckle contrast approaches the level of the noise in the measurement, it is hard to tell if an intensity variation is caused by the signal or the noise.

We illustrated this problem using a 1 m long fiber with spectral correlation width of 0.4 nm. We constructed a transmission matrix with $N = 500$ spatial channels and $M = 500$ spectral channels. The spacing of spectral channels was 0.2 nm, and the bandwidth was 100 nm (from 1450 nm to 1550 nm). In Fig. 6, we examine the ability of the fiber spectrometer to reconstruct Lorentzian shaped spectra with varying width. These probes were synthesized as described above using a weighted sum of sequentially measured speckle patterns. In Fig. 6(a), the probe spectra are shown in red and the reconstructed spectra in blue. While the fiber spectrometer very accurately reconstructs the narrow spectrum, the reconstruction degrades for broader spectra and the reconstruction error increases. The difficulty that the fiber spectrometer has in dealing with broad spectra lies in the reduction of speckle contrast. This can be seen clearly in Fig. 6(b), which shows the speckle patterns corresponding to the three spectra in Fig. 6(a). These speckle patterns are synthesized by the summation of the measured speckle patterns at different wavelengths scaled by the Lorentzian functions with different bandwidth ($\Delta\lambda_L$). Quantitatively, the speckle contrast is calculated as $C = \sigma / \langle I \rangle$, where $\sigma$ is the standard deviation of pixel intensity from the mean value $\langle I \rangle$, and $\langle ... \rangle$ is an average over all pixels across the speckle image. When the input spectrum gets broader, the speckle contrast decreases such that $C \sim 0.1$ for the bandwidth of $\Delta\lambda_L \sim 50$ nm. The number of independent speckle patterns that constitute the speckle image of the Lorentzian signal is approximately $\Delta\lambda_L / \delta\lambda$, and $C$ scales as $[\Delta\lambda_L / \delta\lambda]^{-1/2}$.

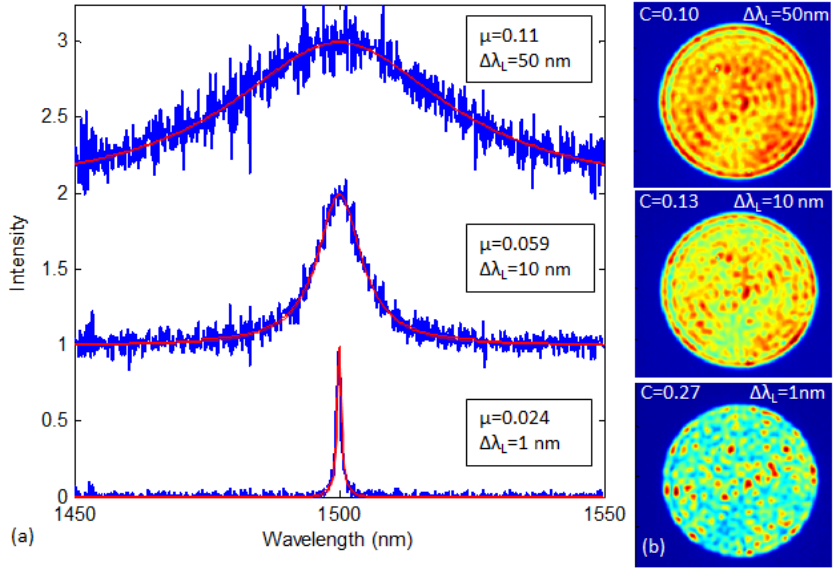

Fig. 6. (a) Reconstructed spectra (blue line) for the probe signals (red line) with Lorentzian spectra of varying width (FHWM) $\Delta\lambda_L$. The spectrum reconstruction error $\mu$ increases with the bandwidth of the probe spectrum. (b) Speckle images corresponding to the three spectra in (a). They are synthesized by summation of sequentially measured speckle patterns where the Lorentzian signals in (a) were used to weight the speckle patterns. The speckle contrast decreases with the bandwidth of the probe signal ($\Delta\lambda_L$), leading to less accurate spectral reconstructions.

The speckle contrast places a practical limit on the bandwidth of the fiber spectrometer in cases when it will be used to measure dense optical spectra. The spectrum reconstruction requires that the speckle contrast be larger than the uncertainty of the experimental measurement. That is: $C \sim [\Delta\lambda/\delta\lambda]^{-1/2} > \sigma_{exp}$ where $\Delta\lambda$ is the operation bandwidth and $\sigma_{exp}$ is the experimental error. We estimated $\sigma_{exp}$ from the standard deviation between two subsequently measured transmission matrices for a 1 m fiber, and obtained $\sigma_{exp} \sim 5\times10^{-3}$. In our implementation, this limits the number of independent spectral channels to a few hundred. Of course, if the spectrometer is used to measure sparse spectra, consisting of only a few narrow spectral lines, this limitation does not apply and much larger bandwidth is possible.

## 7. Polarization-resolved detection

To reduce the spectrum reconstruction error and increase the bandwidth, we shall increase the speckle contrast, which can be done by performing a polarization-resolved measurement. Although the input signal to the multimode fiber had a fixed linear polarization, the polarization after propagating through the fiber was scrambled [14]. Since the two orthogonal polarization components produced distinct speckle patterns, the speckle contrast was reduced even for a single input wavelength. We therefore adapted our experimental setup to separate the two linear polarization components, effectively increasing the speckle contrast by a factor of $\sqrt{2}$. As sketched in Fig. 7(a), two polarizing beam splitters and two mirrors were used to produce spatially displaced images of the speckle patterns with orthogonal linear polarizations on the camera. The inset of Fig. 7(a) shows the speckle images taken at $\lambda = 1500$ nm with the polarizing beam splitters labeled "Polarized Detection" and, as a reference, we removed both beam splitters and compared the speckle image using "Unpolarized Detection". The contrast of the speckle images collected using polarized detection was 0.86, whereas the contrast using unpolarized detection was only 0.6. Ideally a linearly polarized speckle has a contrast of 1, but the finite pixel size of the CCD camera reduces the contrast. The two speckle patterns of orthogonal polarizations taken at the same wavelength are uncorrelated, and their sum gives the unpolarized speckle pattern with a factor of $\sqrt{2}$ lower contrast.

We calibrated transmission matrices using a 1 m long fiber with spectral correlation width of 0.4 nm. The bandwidth was set to 100 nm and the spectral channel spacing was 0.2 nm. In the case of unpolarized detection, we used $N = 2000$ spatial channels. In the case of polarized detection, the number of spatial channels was doubled to take advantage of the second image. Hence the polarization-resolved detection scheme provides two advantages: the speckle contrast is increased by $\sqrt{2}$ and the number of spatial channels is doubled, although the total intensity of each polarized speckle pattern is half of the unpolarized one. In Fig 7(b), we compare spectra reconstructed using the two detection schemes. In both cases, we use a Lorentzian shaped dip in an otherwise flat spectrum as a probe. The probe spectrum is shown in red and the reconstructed spectrum in blue. The polarization-resolved detection clearly provides a more accurate reconstruction, as confirmed by the reconstruction error $\mu$. We repeated this experiment for similar probe spectra with varying Lorentzian widths and plot the spectrum reconstruction error $\mu$ in Fig. 7(c). As the width of the Lorentzian dip decreased the input spectrum became denser, yielding lower speckle contrast, and $\mu$ increased. However, in all cases the polarization-resolved detection gave smaller reconstruction error.

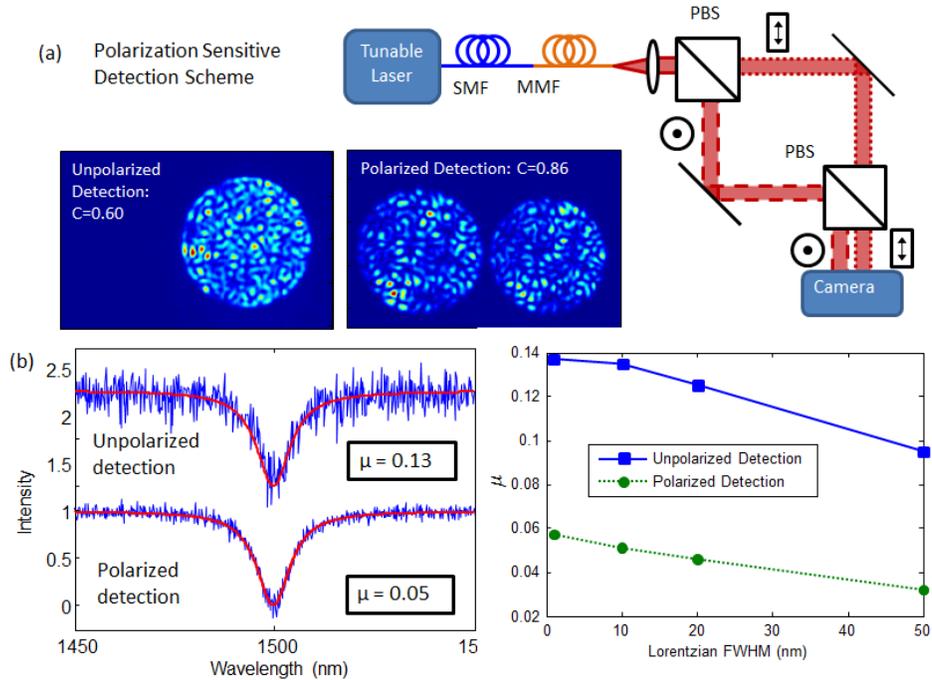

Fig. 7. (a) Schematic of the setup used to separately record the speckle patterns of orthogonal linear polarizations. Two polarizing beam splitters (PBS) and two mirrors are used to separate the two polarization components of the speckle pattern and project both of them to the camera. Right inset: an image of two polarized speckle patterns taken at $\lambda = 1500$ nm. The contrast, $C$, is 0.86, lower than 1 due to the finite pixel size of the CCD camera. Left inset: an unpolarized speckle image taken at the same $\lambda$ with the two polarized beamsplitters removed. The contrast is 0.6, about a factor of $\sqrt{2}$ lower than the polarized speckle, because the speckle patterns of the two polarizations add in intensity. (b) Original (red line) and reconstructed (blue line) flat spectra with a Lorentzian dip. The flat spectrum extended from 1450 nm to 1550 nm, the Lorentizn dip is centered at 1500 nm with a FWHM of 10 nm. The bottom blue curve was obtained with the polarized detection scheme, and the top with unpolarized. The noise in the reconstructed spectra, characterized by the reconstruction error $\mu$, is more than a factor of 2 lower in the polarization-resolved measurement. (c) Error to reconstruct the flat spectrum with a Lorentzian dip of varying width. The broader the Lorentzian dip corresponded to less "dense" spectra, and thus the reconstructed spectra exhibited lower $\mu$. In all cases, the polarized detection scheme (green circles connected by dotted line) provided superior reconstruction than the unpolarized detection (blue squares connected by solid line).

## 8. Measurement of broadband continuous spectra

The speckle patterns used to test the response of the fiber spectrometer in the previous sections were synthesized as linear summations of speckle patterns recorded at different wavelengths using a tunable laser. Here, we demonstrate that the fiber spectrometer can accurately measure the spectra of signals from a broadband source. This also demonstrates that the spectrometer calibration is sufficiently stable to allow us to characterize a different source from the one used to calibrate the transmission matrix. Light from a supercontinuum source (Fianium SC450) was passed through an interference filter and used as the probe signal. At normal incidence, the filter has a ~10 nm wide transmission band centered at 1510 nm. By tilting the filter, the transmission band shifts continuously to shorter wavelength, providing a tunable broadband signal. Since the probe spectra are relatively broad, a fine spectral resolution is not needed, so we used a 2 cm long multimode fiber with a spectral correlation width of 4 nm [Fig. 8(a)]. We adapted the polarization-resolved detection scheme described in the previous section. The fiber transmission matrix was calibrated using 2000 spatial channels and 200 spectral channels equally spaced between 1490 nm to 1530 nm. The filtered supercontinuum emission was then coupled to the input end of the single mode fiber which was connected to the tunable laser during the calibration (see Fig. 1), and a speckle pattern was recorded. The spectrum was reconstructed with the truncated inversion method. An optical spectrum analyzer was used to separately measure the probe spectra, which were compared to the reconstructed spectra of the fiber spectrometer. Figure 8(b) shows three reconstructed spectra (dotted line) centered at 1500 nm, 1505 nm and 1510 nm, they all agree well with the original spectra (solid line). The average spectrum reconstruction error is 0.08. Therefore, the fiber spectrometer accurately recovered the spectra of broadband signals.

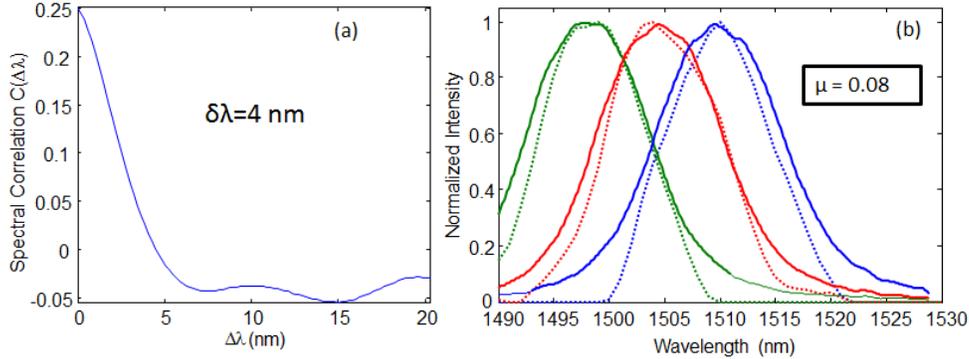

Fig. 8. (a) Spectral correlation function for a 2 cm long multimode fiber. The correlation width $\delta\lambda$ is 4 nm. (b) The 2 cm long multimode fiber was used to measure the spectrum of a supercontinuum source passed through an interference filter. The filter was tilted to provide three probe spectra, centered at 1510, 1505, and 1500 nm. The probe spectra were measured separately using an optical spectrum analyzer (solid lines) and the corresponding spectra reconstructed using the multimode fiber are shown in the dotted lines.

## 9. Discussion and Summary

The studies presented above demonstrate that multimode optical fibers can be used as high-resolution, general purpose spectrometers. The spectral-to-spatial mapping is provided by interference of the guided modes which generates a wavelength dependent speckle pattern. Of course, this speckle pattern is also sensitive to the spatial profile and polarization of the incident optical signal as well as to the environment surrounding the fiber. In fact the speckle of multimode fibers has been used for a variety of environmental sensing applications. For instance, researchers have monitored the speckle to track the changes to the surrounding environmental such as temperature [15], pressure [16], acoustic vibrations [17], or properties of a liquid the fiber was immersed in [18]. The speckle patterns have also been used to sense tiny displacements or vibrations of the fiber position [19-21]. In these environmental sensing applications, the optical probe signal is typically provided by a laser source with a fixed wavelength. In our approach of using the fiber as a spectrometer, the opposite constraint is required: in order to measure changes in the incident optical signal (i.e. wavelength) the environment around the fiber must remain constant.

To provide an estimate of the sensitivity of the multimode fiber spectrometer to changes in the environment, we used the modeling technique presented in Section 3 to study the effect of ambient temperature. For glass fibers, changes in temperature affect both the refractive index, $n$, as $dn/ndT = 7 \times 10^{-6}$ °C$^{-1}$ and the fiber length, $L$, as $dL/LdT = 5 \times 10^{-7}$ °C$^{-1}$ [22]. Using these two expressions, we calculated the speckle patterns at the end of a waveguide with width $W = 1$ mm and NA = 0.22 as a function of temperature (i.e. changes in $n$ and $L$ in our model). We considered waveguides with length $L = 0.1$, 1, or 10 m. We fixed the input wavelength at 1500 nm and calculated the intensity distributions at the output of the waveguide at different temperatures. By correlating the speckle patterns at different temperatures, we obtained the temperature correlation functions, shown in Fig. 8(a). As expected, the correlation function falls off more rapidly for longer fibers. Figure 8(b) presents the temperature correlation widths, defined as twice of the change in temperature required to produce a 50% decorrelation of the speckle pattern. We see that the temperature correlation width scales inversely with the fiber length, confirming that longer fibers are more susceptible to temperature induced changes in the speckle pattern. For a 1 m long fiber, the temperature would need to change by ~8°C to decorrelate the speckle pattern. The effect of the temperature change on the fiber spectrometer performance depends on the spectra it reconstructs, and a detailed study will be presented in the future. One approach to mitigate the temperature sensitivity would be to calibrate a bank of transmission matrices at different temperatures and then use the transmission matrix corresponding to the ambient temperature during a given measurement.

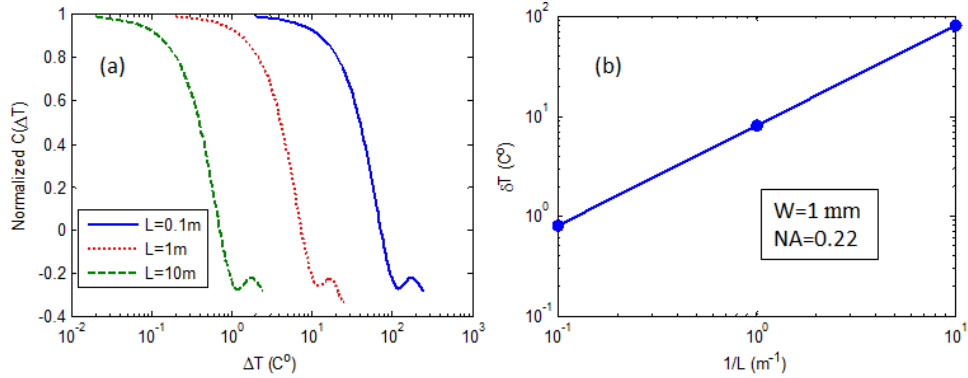

Fig. 9. (a) Calculated temperature correlation function for waveguides with $W$=1 mm, NA=0.22, and $L$=0.1, 1, or 10 m. (b) The calculated temperature correlation width scales inversely with the waveguide length. For $L$=1 m, the temperature correlation width ~8°C.

Fiber-based spectrometers can have much lighter weight, smaller size, and lower cost than traditional grating based spectrometers, while potentially providing state of the art performance in terms of spectral resolution and efficiency (throughput). Large bandwidth can be achieved, especially for the detection of sparse spectra; and, in general, the resolution can be traded for bandwidth, as in a traditional spectrometer. A single multi-mode fiber can operate at varying spectral regions similar to the way traditional spectrometers do by rotating the grating. For a fiber spectrometer, the analog of rotating the grating is achieved by switching the transmission matrix to the one calibrated for the spectral region of interest. Of course, the fiber-based spectrometers also have limitations. They require a calibration step and after the calibration the fiber cannot be twisted or bent further. In addition, the spatial distribution and polarization of the input signal must be identical to that used in the calibration. In our implementation, this was realized by using a polarization maintaining single mode fiber to couple the probe signal to the multimode fiber. If the probe signal has spectral components outside the operating bandwidth, this will introduce additional noise, since the transmission matrix was not calibrated for these wavelengths. Finally, software is required to reconstruct the actual spectrum from the probe signal. Nonetheless, as we showed, the reconstruction algorithm can be fast and robust to experimental noise. The dramatically reduced size and cost of the fiber spectrometers could enable a range of new spectroscopy applications.

**Acknowledgements**

We are grateful to Prof. Hong Tang and Dr. Carlson Schuck for lending us the optical spectrum analyzer. We also thank Profs. Paul Fleury, Allard Mosk, Willem Vos, Doug Stone, and Michael Choma for useful discussions. This work was supported in part by the National Science Foundation under the Grant No. ECCS-1128542.